\documentclass{JHEP3}
\usepackage{graphicx}
\usepackage{bbm}
\usepackage{amssymb,amsmath}

\newcommand{\beq}{\begin{equation}}
\newcommand{\eeq}{\end{equation}}

\newcommand{\rf}[1]{(\ref{#1})} 

\newcommand{\ba}{\begin{array}}
\newcommand{\ea}{\end{array}}

\newcommand{\bracket}[2]{\bra{#1}\,#2\rangle} 
\newcommand{\bra}[1]{\langle\,#1\,|}          
\newcommand{\ket}[1]{|\,#1\,\rangle}          

\newcommand{\p}{\partial} 
\newcommand{\ud}{\mathrm{d}} 

\newcommand{\e}{\mathrm{e}} 
\newcommand{\ad}[1]{\overrightarrow{#1}}
\newcommand{\GS}{Grotowski and Schirmer}
\newcommand{\FP}{Faddeev-Popov}

\title{Colour, copies  and confinement }
\author{Anton Ilderton, Martin Lavelle and David McMullan\\School of Mathematics and Statistics\\
University of Plymouth\\Plymouth, PL4 8AA\\UK\\
E-mail: \email{abilderton@plymouth.ac.uk}\quad
\email{mlavelle@plymouth.ac.uk}\quad
\email{dmcmullan@plymouth.ac.uk}}

\abstract{In this paper we construct a wide class of Gribov copies in Coulomb gauge $\mathop{\rm SU}(2)$ gauge theory. Infinitesimal copies are studied in some detail and their non-perturbative nature is made manifest. As an application it is shown that the copies prevent a non-perturbative definition of colour charge.}
\keywords{Confinement, Non-perturbative Effects, Gauge Symmetry}

\begin{document}

\section{Introduction}
Physical observables are gauge invariant, so it does not seem unreasonable to assume that it should be possible to bypass  the unpleasantness of gauge fixing in any fundamental approach to gauge theories. Indeed, this is one of the virtues of the lattice approach and for important classes of ``gold-plated'' observables, such as  masses or decay constants of stable hadrons, there have been spectacular successes~\cite{Davies:2003ik}. However, not all observables fall into this class and a basic fact of life is that for many important observables, being gauge invariant does not mean that the issues concerning gauge fixing can be avoided. Indeed, important properties of some observables only become clear once their connection to gauge fixing has been clarified. A fundamental example of this, and one that we shall focus on in this paper,  is colour~\cite{Lavelle:1995ty}.

Although it sounds paradoxical, gauge fixing offers a route to the construction of certain gauge invariant quantities. In particular, charges constructed in this way~\cite{Usannals} have better infrared properties than conventional fields and offer a possible resolution to the infrared crisis that plague gauge theories~\cite{Lavelle:2005bt}. The main virtue of this approach to charges is that they are not simply made gauge invariant and then assumed to overlap with some unknown physical state (such as in the use of Wilson lines to probe the inter-quark potential on the lattice) but that their physical significance is built into their very construction from the  beginning.

The process of using gauge fixing to construct a gauge invariant charge is called dressing. Physically, we envisage the matter field being dressed by gauge fields to make it gauge invariant. This, though, is done in a very precise way which reflects the kinematics of the charges involved. So the dressing for, say, a static charge is different to that of a moving charge: each one will have a special gauge fixing that is adapted to it. For a static charge it is the Coulomb gauge that plays the central role and in that gauge the gauge invariant static charge takes its simplest form.

Hence for observables such as colour the thorny  issues associated with gauge fixing take on a new urgency. In particular, the Gribov copies~\cite{Gribov:1977wm} associated with a wide class of gauge fixings cannot be treated as just a technical issue to do with over-counting configurations, but will have direct physical consequences. In~\cite{Lavelle:1995ty}, general arguments were given for the global breakdown of colour charges in QCD due to such copies. This implies that Gribov copies and the underlying global topology of the Yang-Mills configuration space~\cite{Singer:1978dk}, lead inexorably to the absence  of colour in unbroken gauge theories. Many advances are still needed to fill out this route to confinement. The infrared dynamics of these charges is subtle and although there has been some progress~\cite{Usannals,Lavelle:2005bt}, much still needs to be done, particularly in understanding collinear divergences. However, recent lattice simulations~\cite{Langfeld:2007} have used such dressings to calculate the non confining inter-quark potential, which gives additional support to this programme. 

In this paper we want to expand on the arguments given in~\cite{Lavelle:1995ty} and understand precisely how Gribov copies obstruct the construction of a charge in a non-abelian gauge theory. To this end it is desirable to have a set of simple examples of such copies. Although the literature on this topic dates back to Gribov's original 1977 paper (for a recent review see~\cite{Sobreiro:2005ec}), we should note that charges are never considered and that the bulk of such work, see e.g. \cite{Jackiw:1977ng},  deals with gauge transformations $U(x)$ that are non-trivial asymptotically, i.e., $U(x)\nrightarrow \mathbbm{1}$ as $|\underline{x}|\rightarrow\infty$ and as such correspond to \lq large\rq\ gauge transformations which belong to the disconnected part of the group of gauge transformations.

Due to this restriction, simply adding charges to what has been done before is not as simple as it sounds since we have shown~\cite{Lavelle:1995ty} that we must asymptotically have $U(x)\to\mathbbm{1}$ (or at least the centre) for a colour charge to be well defined. Two papers are worth noting in this context. Henyey~\cite{Henyey:1978qd} constructed an explicit set of copies of
an axially  symmetric gauge field, while Grotowski and Schirmer~\cite{Grotowski:1999ay}  proved that a large class of spherically  symmetric configurations have copies. In both cases the gauge transformations are regular ones for which a colour charge can be defined. Although the Grotowski and Schirmer paper gave no examples it is very suggestive as it offers the possibility of a large class of explicit configurations that would allow us to  probe the extent to which colour charges can be defined. It is important to note that in most discussions in the literature the coupling $g$ is hidden. We will include the coupling as this will allow us to demonstrate the non-perturbative nature of the copies even when they are generated infinitesimally. 

The plan of this paper is as follows. After this introduction, in Section~2 we will recap how colour is defined and the significance of this for the allowable gauge transformations. We will then present the dressing approach to the construction of static colour charges which extends the methods given in~\cite{Lavelle:1995ty}. In Section~3 we will enlarge on the arguments given by Grotowski and Schirmer~\cite{Grotowski:1999ay}, and then construct explicit examples of spherically symmetric copies. The role of the centre in these constructions will be also be investigated and this will throw some light on the complicated topology of the Gribov horizons. 

Following this, in Section~4 we will investigate  the extent to which such configurations can have copies within the Gribov horizon. Our results here contradict one of the conclusions in~\cite{Grotowski:1999ay}. We will conclude in Section~5 by tracing the impact of the copies on the colour charges constructed in Section~2. We will see explicitly that the colour acquires a gauge dependence which emerges through the copies. Although this breakdown of colour will be manifestly non-perturbative, we shall see that it can arise even infinitesimally. After a general conclusion we include several appendices where technical results are presented.

\section{A world of colour}
Imagine a world of colour. By this we mean that there exist (asymptotic)  states with well defined colour. The fundamental question at the heart of confinement is then: to what extent is this possible?  Charges in gauge theories are notoriously hard to construct, even in the abelian theory ~\cite{Lavelle:2005bt}. However, we shall see that there is an unavoidable obstruction in unbroken non-abelian theories which stops the partonic, effective colour degrees of freedom from becoming genuine asymptotic states.

To proceed let us fix notation and recall some basic facts about gauge theories. The gauge transformations on the vector potential and matter fields are:
\begin{equation}\label{gauget}
    A_\mu\to A_\mu^U:=U^{-1}A_\mu U+\frac1g U^{-1}\partial_\mu U\qquad \mathrm{and} \qquad \psi\to\psi^U:=U^{-1}\psi\,,
\end{equation}
where $A_\mu=A_\mu^aT^a$ and $T^a$ are anti-hermitian matrices in the appropriate representation of $\mathop{\rm SU}(N)$ and $g$ is the coupling constant. Writing $U=\exp(g\alpha^a(x)T^a)$, we identify the generator of gauge transformations with the charge
\begin{equation}\label{gen}
    G(\alpha)=\int\!\ud^3x\,\, G^a(x)g\alpha^a(x):=\int\!\ud^3x\,\, \Bigl(J_0^a(x)-\frac1g(D_iE_i)^a(x)\Bigr)g\alpha^a(x)\,,
\end{equation}
where $D_i=\p_i+g[A_i,\cdot]$ is the covariant derivative, $J^a_0=-i\bar{\psi}\gamma_0 T^a\psi$ and $E_i^a(x)$ is the chromo-electric field. The Gauss law constraint used to identify physical states is then taken to be\footnote{The relationship between this condition and the gauge invariance of physical states is a subtle one that we do not need to worry about in this paper.}
\begin{equation}\label{gausslaw}
    G^a(x)\ket{\mathrm{physical}}=0\,.
\end{equation}
\subsection{Defining the colour charge}
The conserved Noether  current associated with the gauge transformation is given by
\begin{equation}\label{noether}
    j_\mu^a(x)=J_\mu^a(x)-f^{abc}F^b_{\mu\nu}(x)A^{\nu\,c}(x)\,,
\end{equation}
where $f^{abc}$ are the structure constants. Hence the colour charge $Q=Q^aT^a$ is given by
\begin{equation}\label{colourq}
    Q^a=\int\!\ud^3x\,\,j_0^a(x)=\int\!\ud^3x\,\,\Bigl(J_0^a(x)-f^{abc}A^b_{i}(x)E_i^{c}(x)\Bigr)\,.
\end{equation}
This is conserved but it is not gauge invariant and hence cannot immediately be identified with a physical observable. However, acting on physical states defined by  \rf{gausslaw} it takes a particularly simple form
\begin{equation}\label{qonphys}
    Q\ket{\mathrm{physical}}=\frac1g\int\!\ud^3x\,\, \partial_iE_i(x)\ket{\mathrm{physical}}=\frac1g\lim_{R\to\infty}\int_{S^2_R} \!\ud\boldsymbol{s} \cdot \boldsymbol{E}(x)\ket{\mathrm{physical}}\,,
\end{equation}
where we have used the divergence theorem to get an integral over the spatial two-sphere of radius $R$. In this form it is straightforward to apply a gauge transformation to the charge to get
\begin{equation}\label{qu}
    Q^U=\frac1g\lim_{R\to\infty}\int_{S^2_R} \!\ud\boldsymbol{s} \cdot U^{-1}(x)\boldsymbol{E}(x)U(x)\,.
\end{equation}
If we require a  physical colour charge then we must have $Q^U=Q$ when acting on physical states. To ensure this we must have that  $U(x)$ tends to a constant $U_\infty$ in the centre of the group. So for $\mathop{\rm SU}(2)$ we have a physical colour charge as long as
\begin{equation}\label{su2q}
    U(x)\to \pm\mathbbm{1}\quad \mathrm{as}\quad |\underline{x}|\to\infty\,.
\end{equation}

\subsection{Constructing colour charges}
Having clarified what the colour charge is and the states that it must act upon, we now need to construct physical states that carry colour. An immediate observation is that the matter fields $\psi(x)$ are not suitable since they are not gauge invariant and hence neither can be identified with observables nor have a colour. What is required is a composite operator of the form
\begin{equation}\label{dressq}
    \Psi(x)=h^{-1}(x)\psi(x)\,,
\end{equation}
where the field dependent \emph{dressing} $h^{-1}(x)$ transforms as
 \begin{equation}\label{dresstrans}
    h^{-1}(x)\to h^{-1}U(x) \,,
\end{equation}
under the gauge transformations so as to compensate for the transformation of the matter field. 

\FIGURE{
\includegraphics[width=0.4\textwidth]{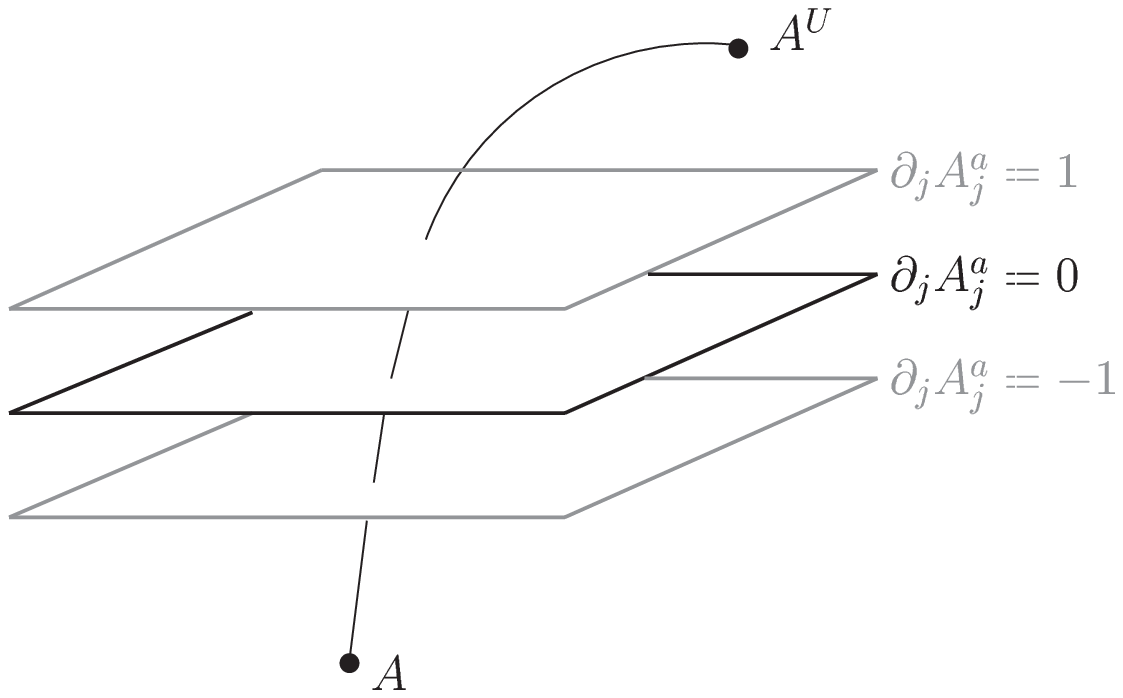}
\caption{The `hoped for' view of gauge fixing -- that the gauge orbits intersect the chosen gauge slice (e.g. $\partial_iA_i^a=0$ in Coulomb gauge) once and only once.
}}

Dressings may be generated using gauge fixing conditions and why this should be so is quite easy to see. Recall that gauge fixing is supposed to pick out a \emph{unique} representative from the orbit of gauge related potentials. So let $\chi(A)=0$ be a gauge fixing condition. For any configuration $A$ there is a \emph{unique} gauge transformation $h\equiv h[A]$ which takes the representative $A$ to the point on the orbit where the gauge fixing condition holds, i.e. $\chi(A^h)=0$. Now consider a different point on the orbit, $A^U$, then there will be a dressing $h^U\equiv h[A^U]$ that takes it to the gauge fixing condition. From this it follows that $\chi((A^U)^{h^U}) = \chi(A^{Uh^U})=0$ and therefore by uniqueness that $h^U=U^{-1}h$ so that $h^{-1}$ transforms as a dressing.

As discussed in~\cite{Lavelle:1999ki}, there are various possible choices for the dressing but most of these would represent highly excited states with little or no overlap with a physical charge. The dressing may be fixed by imposing the condition that the resulting charge is static. In that case, the dressing (or at least the part responsible for making the matter field gauge invariant, which is all which will concern us in this paper, see however,~\cite{Bagan:2001wj}) can be identified with the field dependent gauge transformation which takes a potential $A_\mu$ into Coulomb gauge $\partial_i(A^h)_i=0$. Tests of this construction have been carried out by studying the infrared properties~\cite{Usannals} of the fields and the inter-quark potential~\cite{Lavelle:1998dv,Langfeld:2007} between two such charges.

So, for a static charge we need to solve the Coulomb gauge condition for $h$,
\begin{equation}\label{dressingeq}
   \partial_i(A^h)_i= \partial_i(h^{-1}A_ih+\frac1gh^{-1}\partial_ih)=0\,.
\end{equation}
To help in this we note that the above may be written as
\begin{equation}\label{dressingeq2}
    \partial_i(A_i+\frac1gh^{-1}D_ih)=0\,,
\end{equation}
where now the covariant derivative acts on the group elements by simple commutation: $D_ih=\partial_i+g[A_i,h]$.

If we write $h=e^v$ then, as discussed in Appendix~A, we can write
\begin{equation}\label{abhc1}
    h^{-1}D_ih=\frac{1-e^{\ad{v}}}{\scriptstyle\ad{v}}D_iv\,,
\end{equation}
where we define powers of $v$ acting to the right on some operator $B$ by $\ad{v}^0B=B$, $\ad{v}B=[v,B]$ and $\ad{v}^nB=\ad{v}(\ad{v}^{n-1}B)$. Acting on this by $\partial_i$ we get
\begin{equation}\label{abhc2 }
    \partial_i(h^{-1}D_ih)=\frac{1-e^{\ad{v}}}{\scriptstyle\ad{v}}(\partial_iD_iv)+
    \left[\frac{1-\e^{\ad{v}}}{\scriptstyle{\ad{v}}},\ad{\frac1{\scriptstyle{\ad{v}}}\partial_i v}\right]D_i v\,.
\end{equation}
Substituting this expression into \rf{dressingeq} allows us to construct the static dressing, at least perturbatively. In particular, if we expand $v$ as a power series in the coupling $v=\sum g^nv_{n}$, then we readily find that, for example,
\begin{equation}\label{pdressing2}
    v_{1}=\frac{\partial_iA_i}{\nabla^2}\qquad \mathrm{and}\qquad v_{2}=\frac{\partial_j}{\nabla^2}
    ([v_{1},A_j]+\tfrac12[\partial_jv_{1},v_{1}])\,.
\end{equation}
In this way we can perturbatively identify the static quark with the dressed field
\begin{equation}\label{pertquark}
    \Psi=h^{-1}\psi=\mathrm{e}^{gv_1+g^2v_2+\dots}\psi\,.
\end{equation}
In this and what follows the Laplacian is $\nabla^2=\p_i\p_i$.
\section{Gribov copies}
The simple view depicted in Figure~1 of gauge fixing as selecting a unique representative from each orbit does not hold in practice  and, as shown in Figure~2, the actual interplay between the gauge orbits and the gauge slice can be quite complicated. The existence of such copies, even when the gauge transformations are restricted to lie in the centre asymptotically, are guaranteed by Singer's theorem~\cite{Singer:1978dk}. Here we shall show how to explicitly construct such Gribov copies.
\FIGURE{\centering
\includegraphics[width=0.7\textwidth]{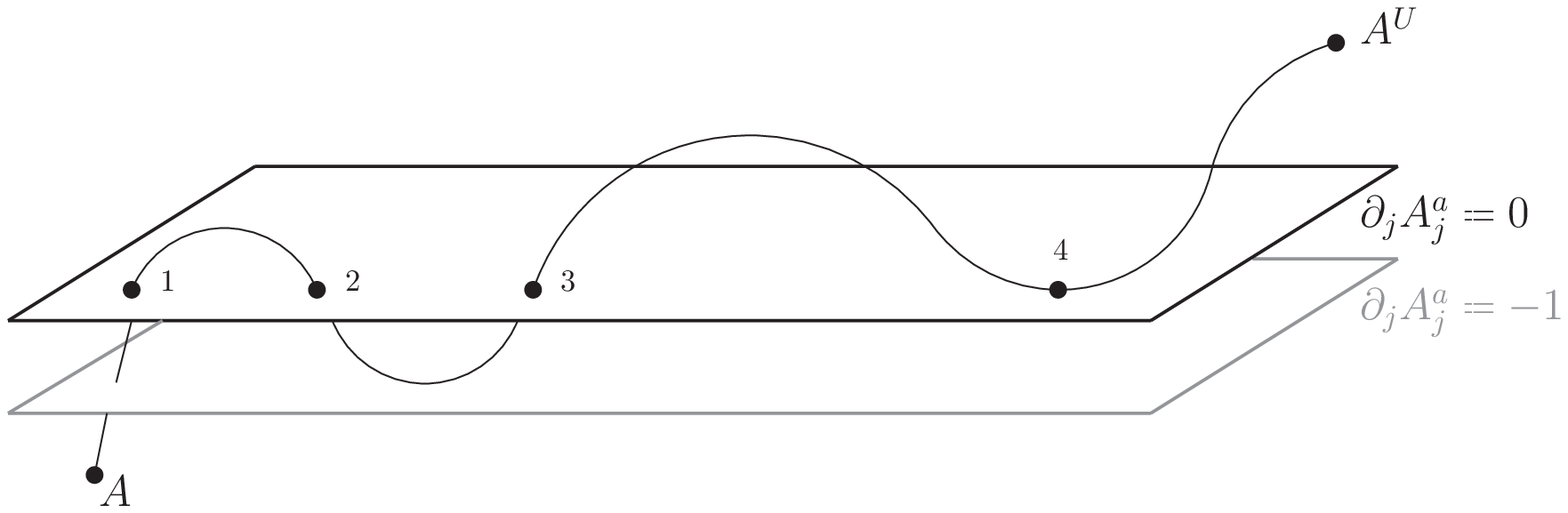}
\caption{The Gribov problem: a gauge field $A$ may be gauge equivalent to multiple copies in the same gauge slice as the gauge orbit $A^U$ intersects (points 1,2,3\ldots) or becomes tangential to  the gauge slice (point 4).
}}

\subsection{Spherical copies}
In the remainder of this paper we will take our gauge group to be SU(2), where
\begin{equation*}
	[T^a,T^b]=\varepsilon^{abc}T^c\,,\qquad T^a=\frac{1}{2i}\sigma^a\,,
\end{equation*}
with $\sigma^a$ the Pauli matrices. We start with configurations of the form
\begin{equation}\label{spha}
A_i^a(x)=\frac{f(r)-1}{r}\varepsilon_{iab}\frac{x^b}{r}\,.
\end{equation}
This is the simplest type of spherically symmetric configuration and is the form considered by \GS\ \cite{Grotowski:1999ay}. The key point to note is that this configuration automatically satisfies the Coulomb gauge condition $\partial_iA_i^a=0$. Consider now the gauge transformation described by
\begin{equation}\begin{split}\label{gt}
  U(x)&=\exp\left(2g\alpha(r)\frac{x^a}{r}T^a\right)\,, \\
    &=\cos{(g\alpha(r))}\mathbbm{1}+i\sin{(g\alpha(r))}\frac{x^a}{r}\sigma^a\,,
\end{split}\end{equation}
where we are setting $\alpha^a(x)=\alpha(r)x^a/r$. This transformation takes $A$ to $A^U$ where
\begin{eqnarray}\label{gta}
(A^U)^a_i&=& \bigg(f(r)\cos(2g\alpha(r))-1 + 2\big(1-g^{-1}\big)\sin^2(g\alpha)\bigg)\frac{1}{r}\varepsilon_{iad}\frac{x^d}{r} \\
\nonumber &&\quad-2\alpha'(r)\frac{x^ix^a}{r^2}-\frac{\sin(2g\alpha)}{r}\big(f(r)-1+g^{-1}\big)\bigg(\delta^{ia}-\frac{x^ix^a}{r^2}\bigg).
\end{eqnarray}
This is an example of a more general spherical configuration. We now impose the condition $\partial_i(A^U)^a_i=0$ to find copies.
Only the final two terms in (\ref{gta}) make a nontrivial contribution to the Coulomb gauge condition, so we need to solve:
\begin{equation}\label{cgta}  \p_i\bigg(2\alpha'(r)\frac{x^ix^a}{r^2}+
\frac{\sin(2g\alpha)}{r}\big(f(r)-1+g^{-1}\big)\bigg(\delta^{ia}-
\frac{x^ix^a}{r^2}\bigg)\bigg)=0.
\end{equation}
From this, using the properties of the transverse projector, we rapidly arrive at the key equation for the existence of copies:
\begin{equation}\label{copies}
  r^2\alpha''(r)+2r\alpha'(r)-\big(f(r)-1+g^{-1}\big)\sin(2g\alpha(r))=0\,.
\end{equation}
This is the equation that we must solve to find the initial configuration, $f$, and the gauge transformation, $\alpha$, which takes us to a copy. Note that \GS\ proved the existence of solutions to this equation (at $g=1$) along with suitable boundary conditions. As such we need to discuss the boundary conditions that we use (which are more general than those considered in~\cite{Grotowski:1999ay}).

For \rf{gt} to be a regular gauge transformation we require that $g\alpha(r)\rightarrow n\pi \ \text{as}\ r\to0\ \text{for}\ n\in\mathbb{Z}$. We may take $n$ to be even or odd so that $U(x)$ approaches $\pm\mathbbm{1}$ at the origin. As we have seen, to define colour we need similar boundary conditions at spatial infinity. Our allowable boundary conditions are therefore
\begin{equation}\label{bcalpha}
  g\alpha(r)\rightarrow \left\{\begin{array}{l} 2n\pi \\ (2n+1)\pi\end{array}\right.\  \text{as}\  r\to0, \infty\implies U(x)\rightarrow \left\{\begin{array}{l} \mathbbm{1} \\ -\mathbbm{1} \end{array}\right..
\end{equation}
Boundary conditions on the function $f$ come from finite energy and norm restrictions. The $L^2$-norm of a configuration $A$ is defined in the obvious way as
\begin{equation}\label{norm}
    \|A\|^2:=\int \!\ud^3x\, A_i^a(x)A_i^a(x)\,.
\end{equation}
Applied to our class of configurations (\ref{spha}) we see that
\begin{equation}\label{spnorm}
    \|A\|^2=8\pi\int_0^\infty\!\ud r\,\,(f(r)-1)^2.
\end{equation}
So if we require a finite norm, given that $f$ is regular, we must have that
\begin{equation}\label{finfty}
    f(r)\to1\ \mathrm{as}\ r\to\infty\,.
\end{equation}
There is no compelling physical reason for imposing such a finite norm condition on our configurations. However, following \GS\ we shall do so here as mathematically it leads to simplifications. Note, though, that weakening this condition is possible and we will discuss this in Section~4.4.  

In contrast to the norm of the configuration, the norm of the field strength is of central importance as it is the energy of the configuration. For our spherically symmetric configuration we see that
\begin{equation}\label{enorm}
\|F_A\|^2=4\pi\int_0^\infty\!\ud r\,\,\frac{(f^2(r)-1)^2}{r^2}+2f'(r)^2.
\end{equation}
Finiteness here implies
\begin{equation}\label{fzero}\begin{split}
    f(r)&\to1\ \text{ as }\ r\to0\,, \\
    f'(r)&\to0\ \text{ as }\  r\to\infty\, .
\end{split}\end{equation}
Note that, as expected, the copy configuration will have the same energy but a different norm given by
\begin{equation}\begin{split}\label{copynorm}
  \|A^U\|^2=4\pi\!\!\int_0^\infty\!\ud r&\,\,2\left[f(r)\cos(2g\alpha(r))-1+2\sin^2(g\alpha(r))(1-g^{-1})\right]^2 \\
  &+ 2\sin^2(2g\alpha(r))(f(r)-1+g^{-1})^2+4r^2(\alpha'(r))^2.
\end{split}\end{equation}
The rate at which the boundary conditions (\ref{bcalpha}), (\ref{finfty}) and (\ref{fzero}) are attained is dictated by the convergence of the norms of $A$, $F_A$ and $A^U$.

\subsection{Explicit examples}
To solve (\ref{copies}) we initially take the boundary conditions used in~\cite{Grotowski:1999ay} where $\alpha\to 0$ for both $r\to0$ and $r\to\infty$. The idea here is to use Henyey's trick \cite{Henyey:1978qd} of inputting the gauge transformation, $\alpha$, by hand and then finding the copied configuration, $f$, via
\begin{equation}\label{f}
f(r)=\frac{r^2\alpha''(r)+2r\alpha'(r)}{\sin(2g\alpha(r))}+1-\frac{1}{g}\,.
\end{equation}
Now comes a simple observation: using the limit $x/\sin(x)\to1$ as $x\to0$ and given the boundary conditions on $\alpha$, we can automatically satisfy the boundary conditions on $f$ if we insist that
\begin{equation}\label{fbca}
    r^2\alpha''(r)+2r\alpha'(r)\to2\alpha(r)\ \mathrm{as}\ r\to0\
    \mathrm{and}\ r\to\infty\,.
\end{equation}
The equation
\begin{equation}\label{larger}
r^2\alpha''(r)+2r\alpha'(r)=2\alpha(r),
\end{equation}
is easy to solve resulting in two types of solution:
\begin{equation}\label{classes}
    \alpha(r)\propto r\qquad\mathrm{and}\qquad \alpha(r)\propto\frac1{r^2}\,.
\end{equation}
Note that these relations need to be interpreted with a little care as $\alpha$ is dimensionless. More properly we should write, for example, that $\alpha(r)\propto r/r_0$ where $r_0$ is some fixed length scale. For simplicity, we shall omit reference to this scale.

Hence we see that if
\begin{equation}\label{alphar}
    \alpha(r)\sim\left\{\begin{array}{ll}
      r & \mathrm{as}\ r\to0\,, \\
      \frac1{r^2} & \mathrm{as}\ r\to\infty\,, \\
    \end{array}\right.
\end{equation}
then we satisfy \emph{both} the boundary conditions on $\alpha(r)$ and those on $f(r)$. These are the only conditions we must satisfy in order to find finite norm, finite energy configurations with Gribov copies, and there are very many ways to solve (\ref{alphar}). The simplest examples are where $\alpha(r)$ is a rational function. For example, take
\begin{equation}\label{ex1}
    \alpha(r)=\frac{r}{1+r^3}\,.
\end{equation}
The resulting $f(r)$ derived from \rf{f} clearly has a coupling dependence. We find
\begin{eqnarray}\label{exf1}
	f(r)=\frac{2r}{(1+r^3)^3}\frac{1-7r^3 +r^6}{\sin\big(\frac{2gr}{1+r^3}\big)}+1-\frac{1}{g}.
\end{eqnarray}
The non-perturbative nature of these solutions is now explicit in the unavoidable presence of factors of $g$ in the denominators. In Figure~\ref{f-examples} we plot two such configurations. \FIGURE{
\includegraphics[width=0.5\textwidth]{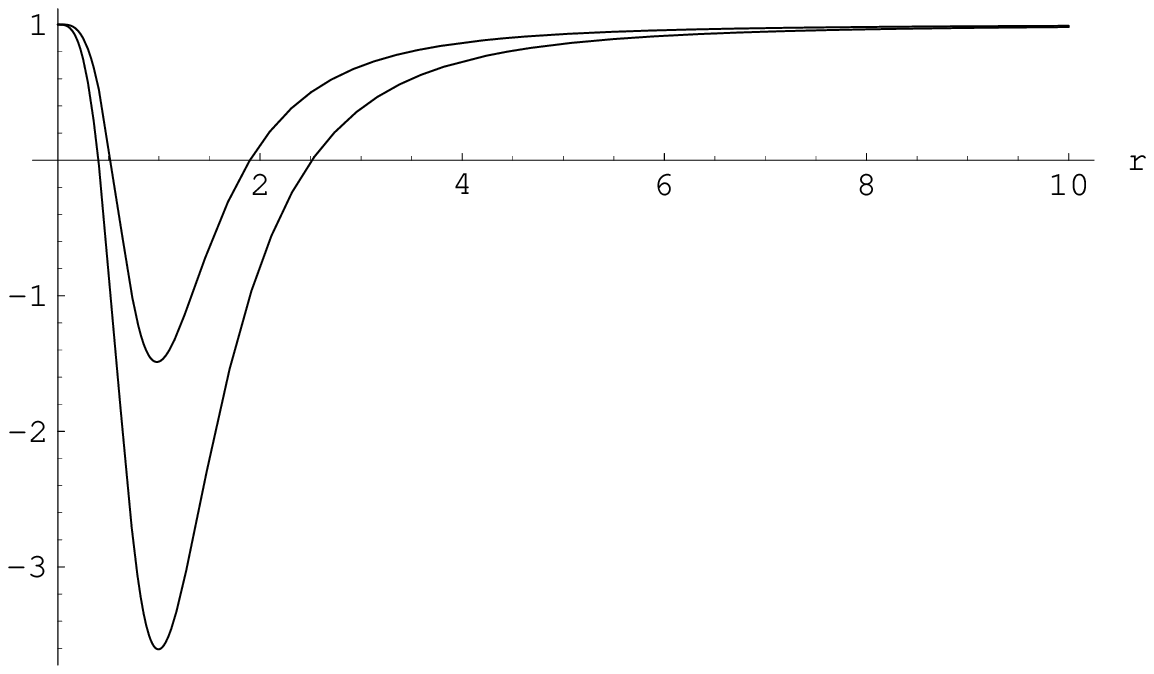}
\caption{\label{f-examples}$f(r)$ as constructed in \rf{ex1} and \rf{exf1}. The upper curve has $g=1$, the lower $g=1/2$.
}}
The norms of this configuration and of its copy, for example at $g=1$, are
\begin{equation*}\begin{split}
	\|A\|&\simeq \sqrt{137}\simeq 11.7\,, \\
	\|A^U\|&\simeq \sqrt{136}\simeq 11.6\,,
\end{split}\end{equation*}
and the energy is $\|F_A\|\simeq \sqrt{340}\simeq 18.4$. Note that the set of connections satisfying the Coulomb condition is a linear subspace of the Yang-Mills configuration space $\mathcal{A}$, which is itself a convex space. Thus we can meaningfully use the norm to measure the difference between configurations. In particular, the distance between the copies is
\begin{equation}\begin{split}\label{dcopies}
\|A^U-A\|^2=4\pi\!\!\int_0^\infty\!\ud r&\,\,2\left[f(r)(\cos(2g\alpha(r))-1)+2\sin^2(g\alpha(r))(1-g^{-1})\right]^2 \\
  &+ 2\sin^2(2g\alpha(r))(f(r)-1+g^{-1})^2+4r^2\alpha'(r)^2.
\end{split}\end{equation}
	Now, following~\cite{vanBaal:1991zw}, if we introduce a parameter $\beta$ via the replacement $\alpha(r)\to\beta\alpha(r)$, then it is clear that
\begin{equation}\label{betazero}
    \|A^{U(\beta)}-A\|^2\to0\ \mathrm{as}\ {\beta\to0}\,.
\end{equation}
That is, the copies coalesce as $\beta$ gets smaller. For $\beta$ very small the two non-perturbative copies are infinitesimally close and we are clearly approaching a configuration on some sort of boundary, a point to which we will return in the next section. Consider the example \FIGURE{
\includegraphics[width=0.5\textwidth]{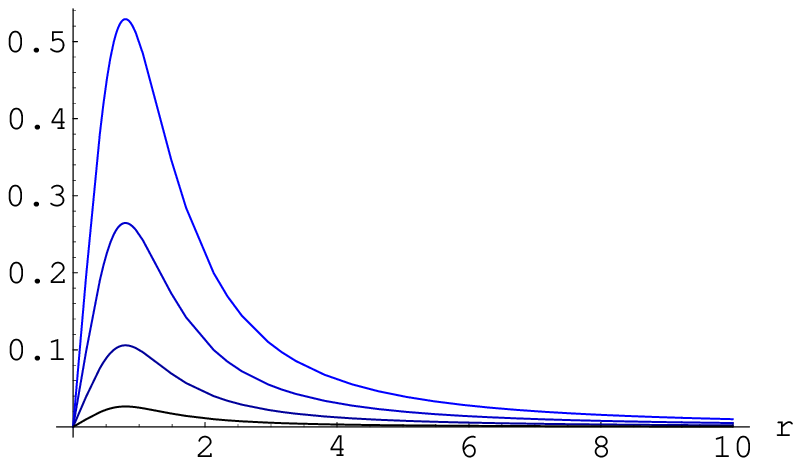}
\label{flow}\caption{As $\beta\rightarrow 0$, $\alpha(r)$ approaches $0$. Here our $\alpha(r)$ is plotted for $\beta=1,1/2,1/5,1/20$.
}}
\begin{equation}\label{limit-trick}
  \alpha(r)=\frac{r\beta}{1+r^3}.
\end{equation}
Provided that $|2g\alpha(r)|<\pi$ (we will return to this point later, for the moment this is easily satisfied, especially as we will shortly take the limit of $\beta\rightarrow 0$) there are no divergences in $f(r)$ and as $\beta\rightarrow 0$ the gauge configuration and its copy coalesce, see Figure~\ref{flow}. The corresponding $f(r)$ in this limit is
\begin{equation}\label{together}
  f(r)=1-\frac{9 r^3}{g(1+r^3)^2}\,,
\end{equation}
with norm $\|A\|=\frac{8\pi}{3^{3/4}g}$. The energy of this configuration is
\begin{equation*}
    \|F_A\|=\frac{32\pi^2(10-21g+24g^2)}{9\sqrt{3}g^4}.
\end{equation*}
\subsection{The centre}
We have so far considered only those transformations with $\alpha(0)=\alpha(\infty)=0$ implying $U(x)$ approaches unity at zero and spatial infinity. Recall, though, that in order to have a regular transformation we must have $g\alpha(r)\to n\pi$ as $r\to0$, so that $U(x)\to\pm1$ there, and that we may relax the boundary condition at infinity to $U(x)\to\pm1$ without losing the ability to define a physical charge.
\TABLE{\begin{tabular}{|l|l|l|}\hline
  & $r\to0$ & $r\to\infty$ \\ \hline
  `Closed' type & $U\to \pm\mathbbm{1}$ &$U\to \pm\mathbbm{1}$ \\ \hline
  `Open' type & $U\to \pm\mathbbm{1}$ & $U\to \mp\mathbbm{1}$ \\ \hline
\end{tabular}\caption{Taxonomy of open and closed boundary types}}
The gauge transformations we consider therefore fall into two classes, best summarised by their boundary conditions, see Table~1. The examples considered previously are of `closed' type -- this language will shortly be clarified. We now wish to give solutions to \rf{copies} for general open and closed gauge transformations.

We begin by expanding the discussion of the limits previously used to solve \rf{copies}. For either open or closed transformations we may write $g\alpha = n\pi + g\omega(r)$ such that $\omega(0)=0$. The periodicity of sine in $f(r)$ leads to the same problem as previously considered,
\begin{equation*}
  f(r) = \frac{r^2\omega''(r)+2r\omega'(r)}{\sin(2g\omega(r))}+1-\frac{1}{g},
\end{equation*}
where $\alpha(r)$ is now replaced by $\omega(r)$. We conclude that $\omega(r)\sim r$ for $r$ small. Now write $g\alpha = m\pi + g\Omega(r)$ such that $\Omega(r)$ vanishes at infinity, then we must solve
\begin{equation*}
  f(r) = \frac{r^2\Omega''(r)+2r\Omega'(r)}{\sin(2g\Omega(r))}+1-\frac{1}{g}
\end{equation*}
and we see that $f(r)$ satisfies the boundary conditions \rf{finfty} if $\Omega(r)$ obeys the same asymptotic behaviour as the original $\alpha(r)$, that is $\Omega(r)\sim 1/r^2$ at infinity. Therefore all our gauge transformations must behave like
\begin{equation*}
	\alpha(r)=n\pi + \text{const}.\,r +\ldots\text{ as }r\to0,\qquad \alpha(r)=m\pi + \text{const}.\,r^{-2} +\ldots\text{ as }r\to\infty.
\end{equation*}
Now that we have identified this shared small $r$ and asymptotic behaviour we will discuss the two classes of transformations in more depth.
\subsection{Open configurations $\alpha_\mathrm{o}$}
`Open' configurations interpolate between $U(0)=\pm\mathbbm{1}$ and $U(\infty)=\mp\mathbbm{1}$. Therefore $\alpha_o(r)$ obeys
\begin{equation}\label{open-behave}
	g\alpha_o(r)\to\begin{cases} 2n\pi, \\ (2n+1)\pi\end{cases}\text{as }r\to 0\quad\text{and}\quad g\alpha_o(r)\to\begin{cases} (2m+1)\pi, \\ 2m\pi\end{cases}\text{as }r\to\infty,
\end{equation}
and clearly must pass through an integer multiple of $\pi/2$ where $f(r)$ diverge unless we construct it carefully. The limits we employ at $r=0$ and $r\to\infty$ may also be applied here. Suppose $g\alpha_o(c)=\pi/2$. Expand $g\alpha_o(r)=\pi/2+g\gamma(r)$ with $\gamma(c)=0$, then
\begin{equation}
  f(r)=-\frac{r^2\gamma''(r)+2r\gamma'(r)}{\sin(2g\gamma(r))}+1-\frac{1}{g}.
\end{equation}
Aside from an initial minus sign, our problem is unchanged from previous arguments. We know that this function will be well behaved at $r=c$ if $\gamma(r)$ is a linear combination of $r$ and $1/r^2$ near $r=c$. The condition $\gamma(c)=0$ implies
\begin{equation}\label{behave}
  \gamma(r)\sim\lambda\bigg(r-\frac{c^3}{r}\bigg)
\end{equation}
near $r=c$ for $\lambda$ some constant. A similar trick works at all multiples of $\pi/2$. This identifies the behaviour of $\alpha_o(r)$ at all potentially problematic interior points.

Given an $f(r)$ and the $\alpha_o(r)$ which generate copies we may consider the gauge transformation $\alpha_o(r/\beta)$ in the limit $\beta\rightarrow 0$. The distance between copies is
\begin{equation*}\begin{split}
\|A^{U(\alpha_o)}-A\|^2=4\pi\!\!\int_0^\infty\!\ud r& \,\,2\left[f(r)(\cos(2g\alpha_o(r\beta^{-1}))-1)+2\sin^2(g\alpha_o)(r\beta^{-1})(1-g^{-1})\right]^2 \\
  &+ 2(f(r)-1+g^{-1})^2\sin^2(2g\alpha_o(r\beta^{-1}))+4r^2\beta^2\alpha'_o(r\beta^{-1})^2.
\end{split}\end{equation*}
Changing variables $r=v\beta$ yields
\begin{equation}\begin{split}
\|A^{U(\alpha_o)}-A\|^2=4\pi\beta\!\!\int_0^\infty\!\ud v& \,\,2\left[f(\beta v) (\cos(2g\alpha_o(v))-1)+2\sin^2(g\alpha_o(v))(1-g^{-1})\right]^2 \\
  &+ 2(f(\beta v)-1+g^{-1})^2\sin^2(2g\alpha_o(v))+4v^2(\p_v\alpha_o(v))^2.
\end{split}\end{equation}
As $\beta\to0$, $f(\beta v)\to1$ for all $v$ and the integral above becomes
\begin{equation*}
\int_0^\infty\!\ud v \,\,2\left[\cos(2g\alpha_o(v))-1+2\sin^2(g\alpha_o(v))(1-g^{-1})\right]^2 + 2g^{-2}\sin^2(2g\alpha_o(v))+4v^2(\p_v\alpha_o(v))^2,
\end{equation*}
which is seen to be finite using the behaviour $g\alpha_o(r)\sim \pi+\text{const}./r^2$ at infinity. As the integral is multiplied by $\beta$ the distance between the copies vanishes as $\beta\to0$.
\FIGURE{\centering
\includegraphics[width=0.5\textwidth]{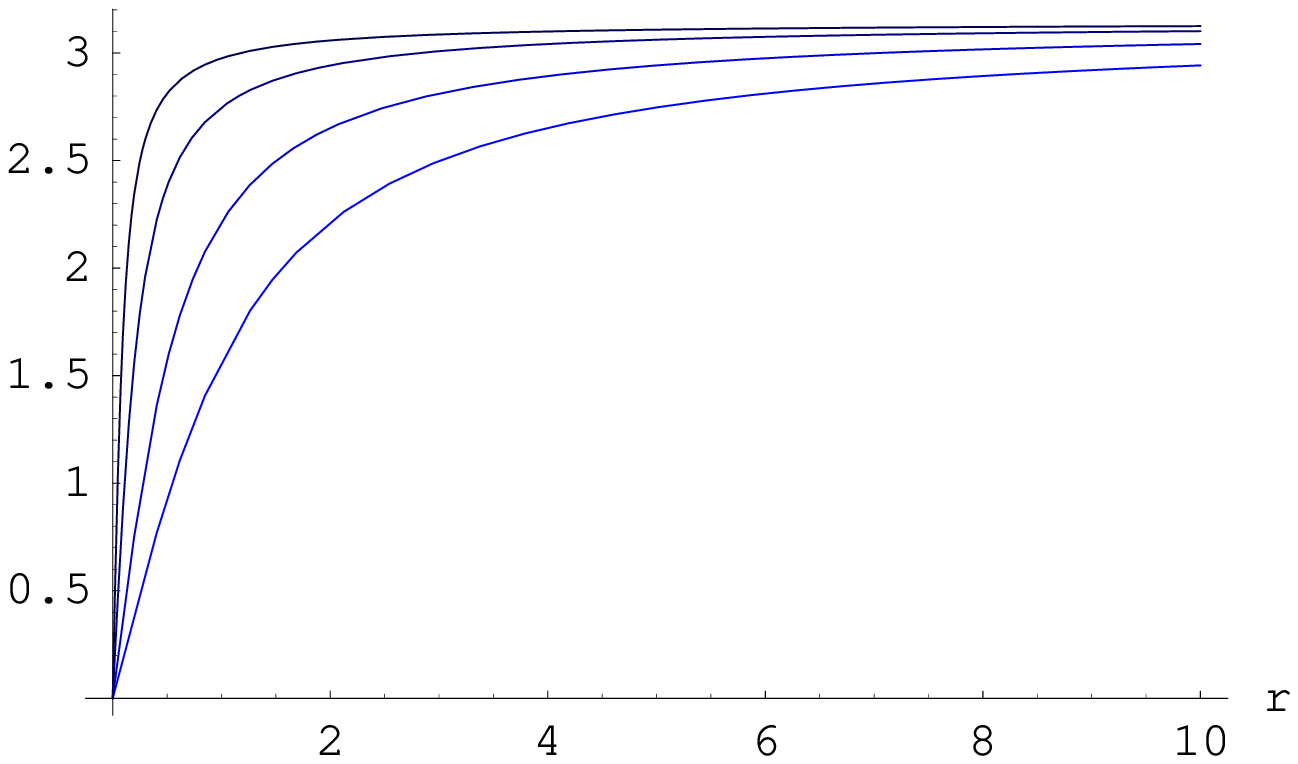}
\caption{\label{flow-open}As $\beta\rightarrow 0$, $g\alpha_o(r/\beta)$ flows to the left, approaching $\pi$ for all non zero values of $r$.
}} In this limit all features of $\alpha$ are localised close to the origin and $U(x)\rightarrow U(\infty)$ in the \emph{open} interval $r\in(0,\infty)$, whereas $U(0)=-U(\infty)$ independent of $\beta$. This is illustrated in Figure~\ref{flow-open} for $\alpha_o(0)=0$, $g\alpha_o(\infty)=\pi$, so that $U(0)=\mathbbm{1}$ independent of $\beta$, but $U(x)\to-\mathbbm{1}$ for all $x\not=0$ as $\beta\to0$.

Compact expressions for open configurations seem hard to find, however in Appendix B we show how they can be constructed by patching functions together.

\subsection{Closed configurations $\alpha_c$}
Closed type configurations $\alpha_c$ interpolate between $U(0)=\pm\mathbbm{1}$ and $U(\infty)=\pm\mathbbm{1}$. When the $\alpha_c$ approach the same multiple of $\pi$ at both zero and infinity they are, in general, given by translations of the $\alpha(r)$ we originally considered (for the special case $\alpha(0)=\alpha(\infty)$), of the form $g\alpha_c(r)=(2n+1)\pi\pm g\alpha(r)$. Provided the original $\alpha(r)$ was suitably bounded so that it did not cause a divergence in $f(r)$ then neither will $\alpha_c(r)$, in fact these result in the same $f(r)$ and copy.

For example, including the parameter $\beta$, an example of a closed transformation is
\begin{equation}\label{ex-closed}
  \alpha_c(r)=\frac{\pi}{g}-\beta\alpha(x)= \frac{\pi}{g}-\frac{\beta r}{1+r^3}.
\end{equation}
This $g\alpha_c$ approaches $\pi$ at both $r=0$ and $r=\infty$ and $f(r)$ is well behaved for all $\beta\leq1$. As $\beta\rightarrow 0$ the gauge transformation approaches $U(0)=U(\infty)$ on the (half) \emph{closed} interval $[0,\infty)$. The distance between copies is, using the translation properties of the trigonometric functions, unchanged,
\begin{equation}\begin{split}
\|A^{U(\alpha_c)}-A\|^2 &=\|A^{U(\alpha)}-A\|^2 \\
&=4\pi\!\!\int_0^\infty\!\ud r \,\,2\left[f(r)(\cos(2g\beta\alpha(r))-1)+2\sin^2(g\beta\alpha(r))(1-g^{-1})\right]^2 \\
  &\hspace{2.1cm}+ 2\sin^2(2g\beta\alpha(r))(f(r)-1+g^{-1})^2+4r^2\beta^2\alpha'(r)^2,
\end{split}\end{equation}
and again the distance between copies vanishes as  $\beta\rightarrow 0$.
\FIGURE{\centering
\includegraphics[width=0.5\textwidth]{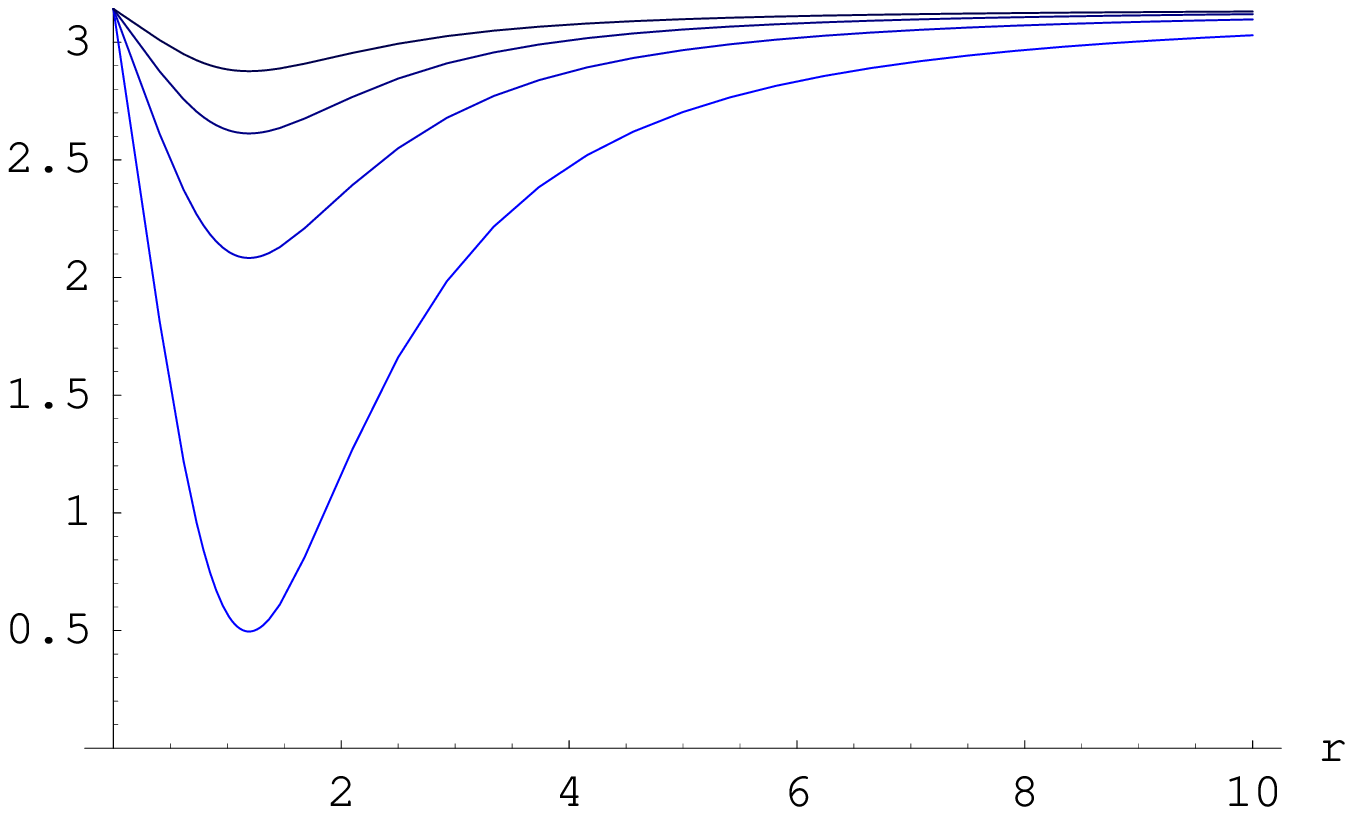}
\caption{\label{flow-closed}The small $\beta$ behaviour of the closed type configuration \rf{ex-closed} with $g=1$. The corresponding gauge transformation $U(x)\to-\mathbbm{1}$ for all $x\in\mathbb{R}^3$ as \hbox{$\beta\to0$}.
}}
If $\alpha_c$ interpolates between different odd multiples of $\pi$ then, provided $\alpha_c$ behaves as in \rf{behave} at multiples of $\pi/2$, we may again consider the transformation $\alpha_c(r/\beta)$ as $\beta\to0$. Here, as in the open case, all features of $\alpha_c(r)$ become localised around $r=0$ and $U(x)$ approaches $U(\infty)$ at all non zero points. The difference between this and the open case is that $U(0)=U(\infty)$ from the outset, so that although $\alpha_c$ only approaches a single value on the open interval $r\in(0,\infty)$, the gauge transformation $U(x)\to U(0)=U(\infty)$ again on the (half) closed interval $r\in[0,\infty)$.

\section{The Gribov horizon}
To clarify what is going on here we need to recall some basic definitions associated with Gribov copies. Given a configuration with its copies, the associated Faddeev-Popov operator is commonly used to pick out a subset of copies.

In the Coulomb gauge, with $\p_iA_i^a=0$ and $U=1+\alpha^a T^a$ an infinitesimal gauge transformation, the equation for the existence of copies becomes
\begin{equation*}\begin{split}
	0&=\p_i\big(U^{-1}A_i U +\frac{1}{g}U^{-1}\p_i U\big) \\
	&=\p_i\big(A_i -[\alpha^aT^a,A_i]+\frac{1}{g}\p_i\alpha^aT^a + \ldots\big) =-\frac{1}{g}\mathcal{D}_A\alpha^aT^a,
\end{split}\end{equation*}
where $\mathcal{D}_A=-\p_iD_i$ is the Faddeev-Popov operator. We see that the condition for the existence of a copy $A^U$ close to the original configuration $A$ is the existence of zero modes in the Faddeev-Popov operator. The first Gribov region is defined as the subset of configuration space where the Faddeev-Popov operator is positive definite. It is bound by the `Gribov horizon', on which the Faddeev-Popov operator picks up its first zero mode. Outside this region, see \cite{Gribov:1977wm} for a fuller discussion, the zero mode becomes a negative eigenvalue mode and further zero modes of the operator appear. We may further decompose configuration space into Gribov regions defined by the number of negative eigenvalue modes of the Faddeev-Popov operator\footnote{Each of these regions is bounded by horizons. It is standard practice to refer to the first region and horizon as `the Gribov region' and `the Gribov horizon', a convention we follow in this paper.}.  Restricting gauge fields to lie within the Gribov horizon is common practice in, for example, lattice calculations (where Coulomb gauge fixing is implemented via a variational argument) but does not eliminate copies, nor is there a physical motivation for this restriction. In this section we will consider the proof in \cite{Grotowski:1999ay} that spherical copies lie inside the Gribov horizon.

\subsection{The Grotowski-Schirmer argument}
We begin with an overview of the \GS\ argument.  The Faddeev-Popov operator in Coulomb gauge is
\begin{equation}
  \mathcal{D}_A = -\p_i D_{i} = -\nabla^2-g[A_i,\p_i\,\cdot\,]\,,
\end{equation}
although for now we restrict to $g=1$, as in \cite{Grotowski:1999ay}. It should be noted that the authors of \cite{Grotowski:1999ay}, to which we will frequently refer, use $\triangle$ for both the full Laplacian and only the radial part in spherical co-ordinates. Given an arbitrary state we may decompose it in spherical vector harmonics. The lowest order term is a radial vector of the form,
\begin{equation}\label{partic}
	\Phi(x)=\phi^a(x)T^a=\phi(r)\frac{x^a}{r}T^a\,,
\end{equation}
on which the Faddeev-Popov operator evaluated at one of our spherical configurations operator takes the simple form
\begin{equation}\label{FP-act}\begin{split}
  (\mathcal{D}_A\Phi)^a &=-\bigg(\phi''+\frac{2}{r}\phi'-\frac{2}{r^2}\phi\bigg)
  \frac{x^a}{r}+2\frac{f(r)-1}{r^2}\phi^a \\
  &=\frac{x^a}{r}\bigg(-\frac{1}{r^2}\p_r r^2\p_r+\frac{2f(r)}{r^2}\bigg)\phi(x).
\end{split}\end{equation}
This expression is positive definite provided $f\geq0$. The \GS\ argument now proceeds by evaluating the effect of the Faddeev-Popov operator on all higher harmonics in the field, and showing that positive definiteness on radial vectors is sufficient to extend the same property to the whole series with only the supplemental constraint $-1\leq f\leq 3$. Already we see a curiosity -- if $f$ can be negative, it no longer seems that \rf{FP-act} may be positive definite (see below). This is easily resolved though, as we simply insist that $0\leq f\leq3$. With this, positivity of $f(r)$ leads to positivity of $\mathcal{D}_A$ which shows that the gauge field (and the copy which may be made infinitesimally close to it, as in \rf{limit-trick}) lie within the Gribov horizon.

\subsection{Reinstating the coupling}
We will now add some detail to the above discussion, along with the factors of $g$. We expand a general field
\begin{equation}
	\Phi(x)=\sum\limits_{jlm}\phi_{jlm}(r)Y^a_{jlm}(\theta,\phi)T^a\,,
\end{equation}
where $Y^a_{jlm}(\theta,\phi)$ are an orthonormal basis for $su(2)$ valued-functions on the 2-sphere carrying orbital angular momentum ($j$) and total angular momentum ($l$). The third component of the total angular momentum is $m$. The sum is understood to be constrained by $|m|\leq l$, $j\geq 0$ and $|j-l|\leq 1$ for $j>0$, $l=1$ for $j=0$. The first term in the series has $j=m=0$, $l=1$ and is the radial field \rf{partic}. The \FP\ operator evaluated at one of our spherical configurations and acting on radial vectors gives
\begin{equation}\begin{split}
	(\mathcal{D}_A\Phi)^a &= -\bigg(\phi''_{010}+\frac{2}{r}\phi'_{010}-\frac{2}{r^2}\phi_{010}\bigg)\frac{x^a}{r}+2g\frac{f-1}{r^2}\phi_{010}\frac{x^a}{r} \\
	&=\frac{x^a}{r}\bigg(-\frac{1}{r^2}\p_r r^2\p_r + 2g\frac{f-1+g^{-1}}{r^2}\bigg)\phi_{010}.
\end{split}\end{equation}
The matrix element of the Faddeev-Popov operator between such fields is given by
\begin{equation}
	\int\!\ud^3x\,\,\phi^a(\mathcal{D}_A\phi)^a = 4\pi\int\limits_0^\infty\!\ud r\, r^2\,\,\phi_{010}\bigg(-\frac{1}{r^2}\p_r r^2\p_r + 2g\frac{f-1+g^{-1}}{r^2}\bigg)\phi_{010}.
\end{equation}
The sign of this expression determines the positivity of the Faddeev-Popov operator. The radial Laplacian is a positive definite operator (for suitable boundary conditions), so the sign depends crucially on the $f$-dependent multiplicative term. Since we may in principle always find a field for which this matrix element is dominated by contributions from the multiplicative term, if $f-1+g^{-1}$ becomes negative the matrix element also becomes negative. For example, take $g=1$ and $\alpha(r)$ again as in \rf{ex1}. The corresponding $f$ is given in Figure~\ref{f-examples} and becomes negative. Now choose $\phi_{010}(r)=\alpha(r)$. The field $\Phi(x)=\phi_{010}(r)x^a/r$ has finite norm,
\begin{equation*}
	\|\phi\|^2=\int\!\ud^3x\,\,\phi^a(x)\phi^a(x) = 4\pi\int\limits_0^\infty\!\ud r\, r^2\alpha(r)^2=\frac{4\pi}{9\sqrt{3}}.
\end{equation*}
The matrix element of the Faddeev-Popov operator acting on this $\phi$ is
\begin{equation*}\begin{split}
  \bra{\Phi}\hat{\mathcal{D}}_A\ket{\Phi}&=4\pi\int\limits_0^\infty\!\ud r\, r^2\,\,\alpha(r)\bigg(\frac{2f(r)}{r^2}-\frac{1}{r^2}\p_r r^2\p_r\bigg)\alpha(r) \\
  &= 4\pi\int\limits_0^\infty\!\ud r\, r^2\,\,\alpha(r)\bigg(\frac{2f(r)\alpha(r)}{r^2}-
  \frac{f(r)\sin(2\alpha(r))}{r^2}\bigg) \\
  &\simeq -0.85.
\end{split}\end{equation*}
and a plot of the integrand is given in Figure~\ref{integrand}. Thus the Faddeev-Popov operator is not positive definite.
\FIGURE{\centering
\includegraphics[width=0.5\textwidth]{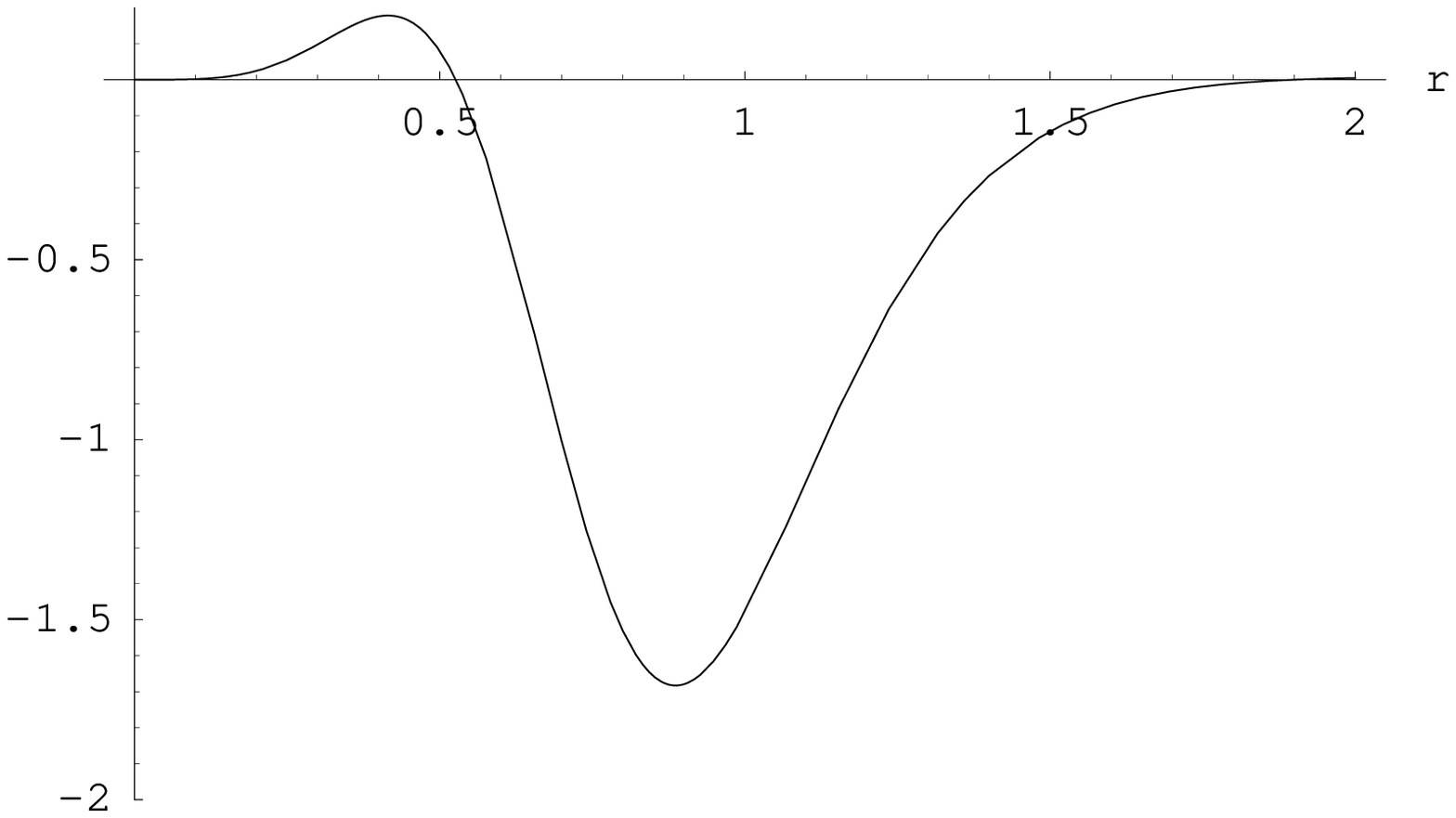}
\caption{\label{integrand}The density of the matrix element $\bra{\Phi}\mathcal{\hat{D}}_A\ket{\Phi}$.
}} We have now seen that the positivity of the lowest term in the harmonic expansion depends on the sign of the combination
\begin{equation}
	\bar f(r):=g\bigg(f(r)-1+\frac{1}{g}\bigg)\,.
\end{equation}
We now turn to the higher harmonics. Writing $A_i^a(x)$ in terms of $\bar f$, the Faddeev-Popov operator becomes
\begin{equation*}
	\mathcal{D}_A = -\nabla^2-[\bar A,\p_i\cdot]\,,
\end{equation*}
where the gauge field $\bar A$ is defined by
\begin{equation*}
	\bar A_i^a(x):=\frac{\bar f(r)-1}{r}\varepsilon_{iab}\frac{x^b}{r} = gA_i^a(x).
\end{equation*}
All the factors of the coupling have been tidied away into $\bar f$. The remainder of the \GS\ argument now follows with $f(r)$ in \cite{Grotowski:1999ay} replaced by $\bar f$. So, we divide the matrix element $\bra{\Phi}\mathcal{D}_A\ket{\Phi}$
into four pieces, I, II, III and IV. The first contains the $j=0$ term which is given above, the remainder correspond respectively to $j\geq 1$ with $l=j$, $l=j-1$, $l=j+1$. The second term is
\begin{equation}
	\text{II}=4\pi\sum\limits_{m,\,l=j}\int\limits_0^\infty\!\ud r r^2\,\,\phi_{llm}(r)\bigg(-\frac{1}{r^2}\p_r r^2\p_r + \frac{l(l+1)+\bar f(r)-1}{r^2}\bigg)\phi_{llm}(r).
\end{equation}
In order for II to be positive we require $\bar f(r)\geq -1$. However we have already seen that $\bar f\geq0$ is required for positivity in the radial part. The third term is
\begin{equation}
	\text{III}=4\pi\sum\limits_{m,\,l=j-1}\int\limits_0^\infty\!\ud r r^2\,\,\phi_{jlm}(r)\bigg(-\frac{1}{r^2}\p_r r^2\p_r + \frac{l(l+2-\bar f(r))}{r^2}\bigg)\phi_{jlm}(r)\,,
\end{equation}
and since $j\geq 2$ (for $j=1$ the last term vanishes) we require $\bar f\leq 3$. The final term is
\begin{equation}
	\text{IV}=4\pi\sum\limits_{m,\,l=j+1}\int\limits_0^\infty\!\ud r r^2\,\,\phi_{jlm}(r)\bigg(-\frac{1}{r^2}\p_r r^2\p_r + \frac{(l+1)(l-1+\bar f(r))}{r^2}\bigg)\phi_{jlm}(r)\,,
\end{equation}
which is non-negative as $l\geq2$ here. Therefore, for $0\leq \bar f\leq 3$ each term in the Faddeev-Popov matrix element is positive, giving us a positive definite operator. This is the extension of the \GS\ argument to arbitrary $g$.

\subsection{Outside the horizon}
As we have demonstrated, if $\bar f$ becomes negative then we may find fields (we gave a radial field as an example) such that the Faddeev-Popov operator fails to be positive definite. In this section we will show that all $\bar f$ must somewhere become negative (for any $g$), and therefore the corresponding gauge fields must lie outside the Gribov horizon.

Let $f(r)$ and $\alpha(r)$ be given by
\begin{equation}\begin{split}\label{prf-cond}
  f(r)&=\frac{r^2\alpha''(r)+2r\alpha'(r)}{\sin(2g\alpha(r))}+1-1/g, \\
  \implies \bar f(r)&=\frac{gr^2\alpha''(r)+2rg\alpha'(r)}{\sin(2g\alpha(r))} \\
  g\alpha(r) &\sim \begin{cases} n\pi + \lambda r\text{ as }r\to0, \\ m\pi + \mu/ r^2\text{ as }r\to\infty.\end{cases}
\end{split}\end{equation}
We will now prove that all such $\bar f(r)$ must take negative values. To begin, if $g\alpha(r)$ interpolates between different multiples of $\pi$ at $0$ and $\infty$ then at some point it must cross an odd multiple of $\pi/2$. We have already seen that in order for $\bar f(r)$ to be well behaved at such a point, call it $c$, then we must have
\begin{equation*}\begin{split}
  g\alpha(r)&=(2n+1)\frac{\pi}{2} + g\gamma(r), \\
  \gamma(r)&=\text{const.}\bigg(r-\frac{c^3}{r^2}\bigg)+\ldots\text{ near $r=c$}.
\end{split}\end{equation*}
At such a point, $\bar f(c)$ is given by
\begin{equation*}\begin{split}
  \bar f(c)&=\lim_{r\to c}\,\, \frac{r^2g\gamma''(r)+2rg\gamma'(r)}{\sin((2n+1)\pi+2g\gamma(r))} = \lim_{r\to c}\,\, -\frac{r^2g\gamma''(r)+2rg\gamma'(r)}{\sin(2g\gamma(r))} \\
  &= -1.
\end{split}\end{equation*}
Thus $\bar f(r)$ must take negative values in a region around $r=c$. Taking, for example, $g=c=1$, the resulting $\bar f(r)$ as $\alpha(r)$ passes through $\pi/2$ at $r=1$ is given in Figure~\ref{midlimit}. \FIGURE{\centering
\includegraphics[width=0.5\textwidth]{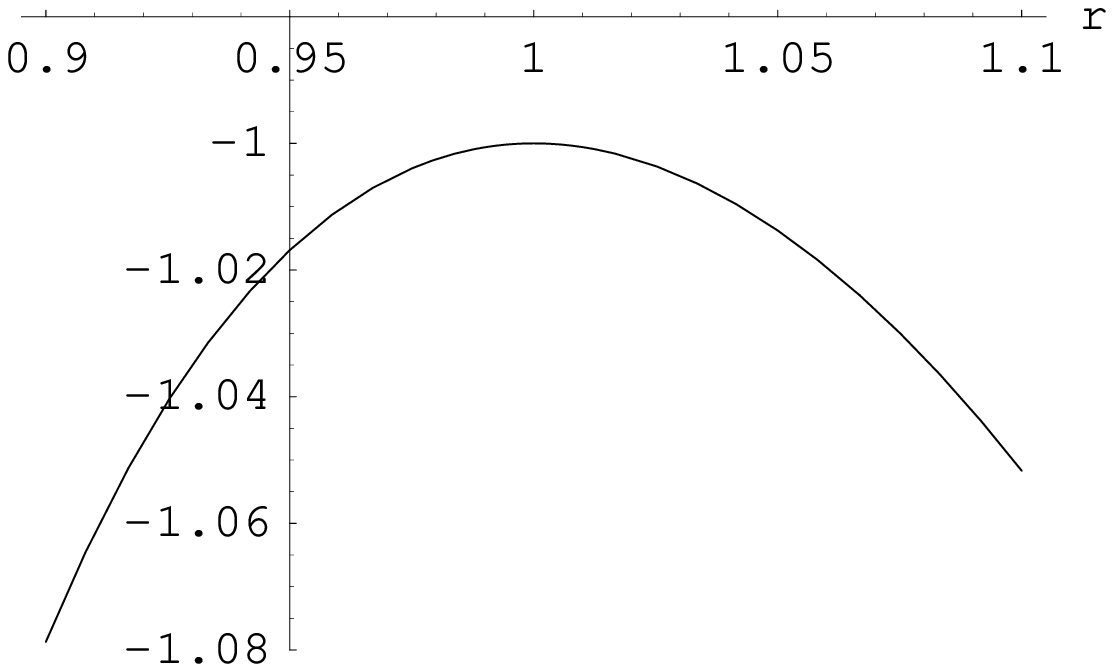}
\caption{\label{midlimit}Example of the behaviour of $f(r)$ at points (here $r=1$) where $g\alpha$ becomes a multiple of $\pi/2$}}
We may now restrict to $\alpha(r)$ which take the same value at $r=0$ and $r=\infty$. Expand $g\alpha(r)=m\pi+g\gamma(r)$ where $\gamma(r)$ vanishes at $r=0$, $r=\infty$. The corresponding gauge configurations are
\begin{equation}
  \bar f(r)=\frac{r^2g\gamma''(r)+2rg\gamma'(r)}{\sin(2g\gamma(r))}, \\
\end{equation}
which is independent of $m$. All such gauge transformations lead to the same configurations and copies. If these $g\alpha(r)$ cross (multiple times) the value $\pi/2$ then they must behave as above and become negative. Therefore we need consider only the case where $\alpha(r)$ obeys
\begin{equation}
  \alpha(0)=0,\qquad \alpha(\infty)=0,\qquad |g\alpha(r)|<\pi/2.
\end{equation}
An example of this kind is given by our staple, \rf{ex1}. We know that for such $\alpha(r)$ when $r$ is small $\alpha(r)\sim ar$ with the constant $a$ free. When $a>0$ there must exist a point $r^*$ such that $\alpha(r^*)$ is a positive maximum; this follows from the behaviour at large r where $\alpha$ behaves like $b/r^2$ and demanding that $\alpha$ be continuous. At the turning point
\begin{equation}\label{fr*}
  \bar f(r^*)=\frac{{r^*}^2g\alpha''(r^*)}{\sin(2g\alpha(r^*))}.
\end{equation}
The second derivative is negative as this is a maximum. Since $\alpha(r^*)>0$ and by assumption $0<2g\alpha<\pi$ this implies $\sin(2g\alpha)>0$. By continuity $\bar f(r)$ is then negative at and in a region around $r=r^*$.

If at small $r$, $\alpha(r)\sim ar$ with $a<0$ then there must exist an $r^*$ such that $\alpha(r^*)$ is a negative minimum. In this case the numerator of (\ref{fr*}) is positive but the denominator is negative and $\bar f(r^*)$ is again negative.

This completes the proof that the $\bar f(r)$ we construct always take (both positive and) negative values. One may imagine that there are other ways to take the limit \rf{fbca}, for example by adding higher powers of $r$ or $\alpha(r)$ which vanish in the limit. Such terms can not affect the leading order behaviour of $\alpha(r)$ as $r\to0$ and $\infty$, however, as these are necessary for the limit to exist and $\bar f(r)$ to be well defined. As it is only this leading order behaviour which we used to prove that the $\bar f(r)$ become negative, such corrections do not affect our argument.

In summary, we have seen that no spherical copies constructed in this way lie within the Gribov horizon, contrary to the statements made by \GS, as the Faddeev-Popov operator at such configurations fails to be positive definite when acting on radial vectors.

\subsection{Monopole configurations}
As mentioned earlier, it is not physically required that our configurations have finite norm, only finite energy. Without the restriction of finite norm we may generate copies of configurations which behave asymptotically as monopole solutions (we are here thinking of embedding our configurations in a Georgi-Glashow type model). We begin with the spherically symmetric configuration 
\begin{equation}
	A_i^a(x)=\frac{\bar{f}(r)-1}{gr}\epsilon_{iab}\frac{x^b}{r},
\end{equation}
which we have written in terms of $\bar{f}(r)$. We will ask that $\alpha(r)$ and $\bar{f}(r)$ behave as before for small $r$, but that $\bar{f}(r)\to0$ as $r\to\infty$ so that we asymptotically approach the usual monopole profile. This change in boundary condition requires a change in the asymptotic behaviour of $\alpha(r)$. It is not difficult to show that if $g\alpha(r)=c/r +g\bar{\alpha}(r)$, where $\bar{\alpha}(r)$ behaves as $1/r^2$ and higher terms as before, that the asymptotic behaviour of $\bar{f}(r)$ becomes
\begin{equation}\begin{split}
	\bar{f}(r)&=\frac{r^2g\alpha''+2rg\alpha'}{\sin(2g\alpha)} \\
	&\sim\frac{2g\bar\alpha}{\sin(2g\bar\alpha)}\times\mathcal{O}\bigg(\frac{1}{r}\bigg),
\end{split}\end{equation}
which vanishes as $r\to\infty$ as required. \FIGURE{\centering
\includegraphics[width=0.5\textwidth]{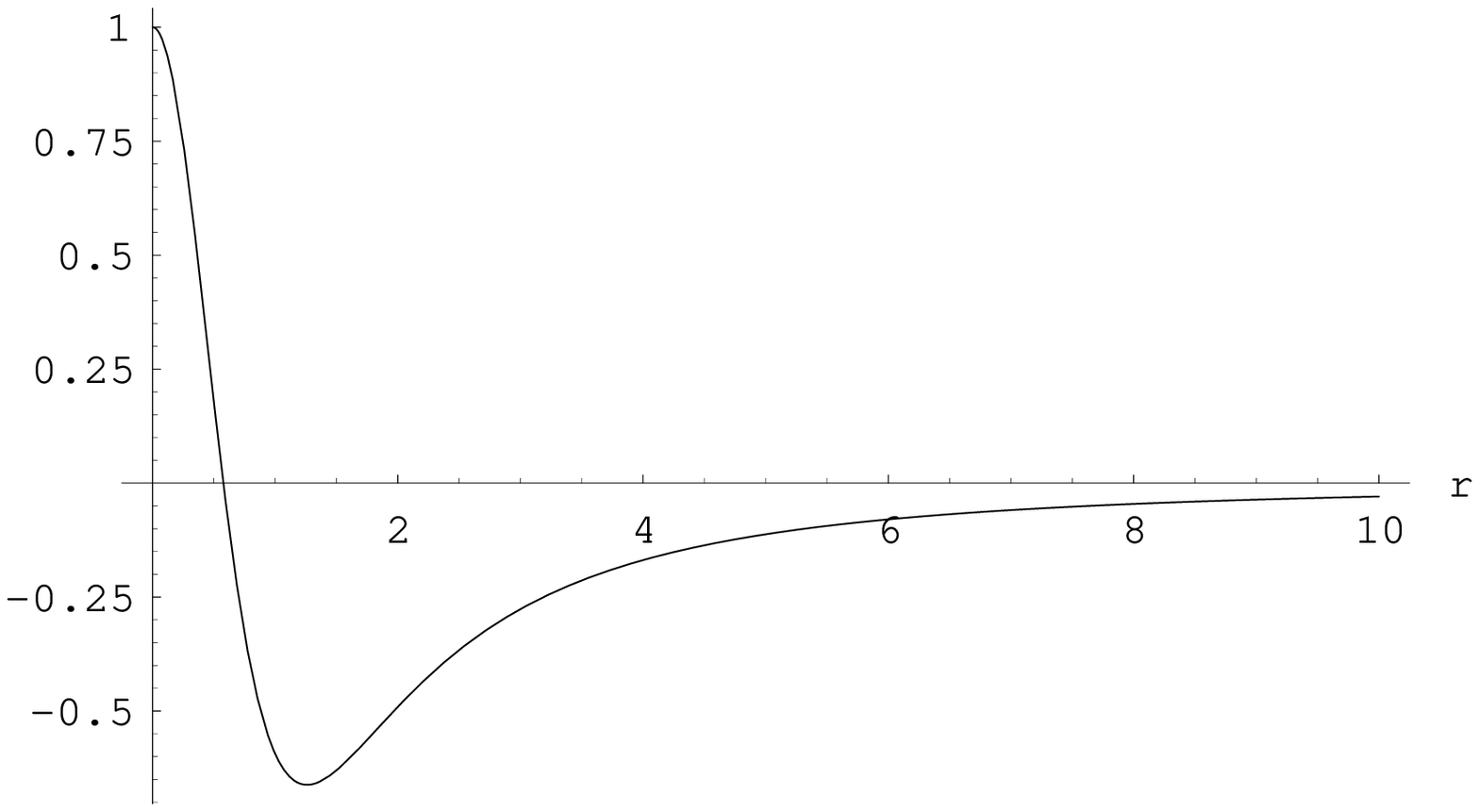}
\label{monopole}\caption{The gauge field $\bar{f}(r)$ constructed from \rf{mono-alpha}, with $g=1$, which behaves asymptotically as a monopole and has Gribov copies.}} Note that such configurations, like those found above, have finite energy. An example is to take
\begin{equation}\label{mono-alpha}
	\alpha(r)=\frac{r}{1+r^2},
\end{equation}
which at $g=1$ has energy $\|F_A\|\simeq 10.6$. The corresponding gauge configuration is shown in Figure~\ref{monopole}. All our earlier arguments extend immediately to this class of configurations; the copies may be made infinitesimally close, $\bar{f}(r)$ must take negative values and so such gauge fields lie outside the Gribov horizon.

\section{The loss of colour}
%
%
%
%

We now return to the construction of colour charges started in Section~2. Recall that the ability to associate colour with the charge~\rf{dressq} relied on it being gauge invariant. This followed from the transformation properties of the dressing~\rf{dresstrans} which in turn was derived from the uniqueness of the Coulomb gauge fixing. The fact that we have copies will invalidate this argument. In this section we want to make this breakdown of colour explicit. 

A question not clarified in Section~2 is: what colour does the charge~\rf{dressq} have? At first sight this looks quite complicated as $\Psi(x)$ is a composite operator constructed out of both the matter and gauge fields which can both contribute to colour. However, the overall colour is quite simple to determine in practice. Before we do this, we should say that in the bulk of this paper we have described $A(x)$ in terms of a classical field, rather than a quantum operator. We are thinking of such objects, and the matter fields, as the eigenvalues in the Schr\"odinger representation. Wave functionals $\zeta[A_i,\psi]=\bracket{A_i,\psi}{\zeta}$ represent quantum states $\ket{\zeta}$ on which $\hat{A}_i(x)$ and $\hat\psi(x)$ are diagonalised. We satisfy the canonical commutation relations by representing the conjugate momenta $\hat{E}^a_i$ and ${\hat{\psi}}^\dagger$ by functional differentiation. The dressed operator $h^{-1}\psi$ is therefore also diagonalised. In this way we may perform the manipulations below, understanding them to represent the action of operators on wave functionals.

Neglecting copies for the moment, gauge invariance of  $\Psi(x)$ implies that its colour is constant on any gauge orbit. Making this more explicit, we write 
\begin{equation}\label{qapsi}
    \Psi[A,\psi](x)=h^{-1}[A](x)\psi(x)\,,
\end{equation}
then gauge invariance means
\begin{equation}\label{gqapsi}
    \Psi[A^U,\psi^U](x)=h^{-1}[A^U](x)\psi^U(x)=h^{-1}[A](x)UU^{-1}\psi(x) =\Psi[A,\psi](x)\,.
\end{equation}
On any orbit there exists a configuration where the Coulomb gauge condition holds and hence the static dressing becomes the identity. At this point the colour charge is just the matter field and we can easily arrange that it is, say, a localised  blue charge there. Anywhere else on the orbit the dressing will not be trivial but the overall colour will still be blue. In this sense, the dressing is transparent to the matter's colour. As we make a gauge transformation away from the Coulomb gauge configuration the matter starts to lose its colour but the dressing compensates so that the overall colour is preserved. That is, \emph{as long as the dressing is sensitive to the applied gauge transformation.}

The existence of copies, though, means that we have gauge transformations for which the dressing is insensitive. In particular, the dressing evaluated on the configuration \rf{spha} is unity as the gauge field is already in Coulomb gauge,
\begin{equation}
	\Psi[A,\psi](x)=h^{-1}[A](x)\psi(x)=\psi(x).
\end{equation}
Performing the gauge transformation \rf{gt} parameterised by $\alpha(r)$ obeying \rf{alphar}, the gauge field is transformed to its copy \rf{gta}, which is also in Coulomb gauge and so the dressing remains unity. This does not compensate for the gauge rotation of the matter field generated by \rf{gt}, so that
\begin{equation}
	\Psi[A^U,\psi^U](x)=U^{-1}\psi(x)=\cos{(g\alpha(r))}\psi(x)-i\sin{(g\alpha(r))}\frac{x^a}{r}\sigma^a\,\psi(x).
\end{equation}
It is clear that the commutator of this transformed $\Psi$ with the colour charge operator will not be the same as that of the original $\Psi$, which has therefore lost its colour.

The copies may be made infinitesimally close together, yet this breakdown of colour is non-perturbative, as demonstrated by the coupling dependence of our explicit examples. However, it should be possible to develop a perturbation expansion around such a solution and see the breakdown of colour in perturbation theory. As a starting point for this and other investigations there are other classes of configurations in the literature where non-perturbative effects becomes apparent upon the inclusion of the coupling. We illustrate this in Appendix C with Henyey's axially symmetric solutions.

\section{Conclusions}
There are two main conclusions from this paper. The first is that explicit examples of Gribov copies in Coulomb gauge are readily produced. These copies arise not only for regular gauge transformations that are smoothly connected to either the identity mapping or the centre of the gauge group, but also for regular transformations with mixed asymptotic conditions which we saw generated open boundaries. Let us re-emphasise that this is a wide class of copies -- for the case of small gauge transformations {\it any} suitably bounded function $\alpha(r)$ with the small and large $r$ behaviour
\begin{equation*}
	\alpha(r)\sim r\quad\text{as}\quad r\to 0\qquad\text{and}\qquad \alpha(r)\sim \frac{1}{r^2}\quad\text{as}\quad r\to \infty
\end{equation*}
gives a finite norm, finite energy configuration and copy as in \rf{spha}, \rf{gta}, \rf{f}. Although the copies may be infinitesimally close to each other the configurations are non-perturbative. This point was only apparent after inclusion of the coupling, which we feel it is important to maintain in future studies.

The other major conclusion from this analysis is the impact these copies have on the construction of colour charges and confinement. Although one may define colour perturbatively, such a definition breaks down non-perturbatively. We constructed a matter field on which a physical colour was well defined provided that the gauge fixing condition, used in the dressing, picked out a unique representative on each gauge orbit. It is here that the copies enter, the existence of which contradicts this uniqueness and gives the colour a gauge dependence. This shows that coloured charges cannot correspond to physical asymptotic sates and are therefore confined to colourless bound states. This has disentangled the fundamental reason for confinement from the detailed dynamics which dictate the scale of hadronisation.

An immediate open question is how we may extend our methods to create copies which lie inside the Gribov horizon. An obvious choice is to change the form of the gauge fields, as in Henyey's class of copies~\cite{Henyey:1978qd} where spherical symmetry is replaced by axial symmetry. Alternatively, we could change the form of the gauge transformation. 

A deeper question to be addressed is that of the physical significance of the Gribov horizon. To what extent does it matter whether a given configuration lies inside or outside the horizon? There is at present no physical motivation for working inside the horizon -- such a choice is essentially arbitrary as we could choose to work in any other Gribov region, nor does the choice eliminate the problem of copies. We feel that this work and such future investigations as outlined above will help towards understanding these important topics.

Note added in proof: another method to try to avoid gauge fixing is the gauge invariant exact renormalisation group, see e.g. \cite{Arnone:2005fb}.
\acknowledgments
It is a pleasure to thank Tom Heinzl, Paul Jameson, Arsen Khvedelidze and Kurt Langfeld for useful discussions. In the early stages of this work one of us (DM) benefited from very helpful conversations with Shogo Tanimura on Gribov copies and correspondence with Joseph Grotowski.

\appendix

\section{Baker-Campbell-Hausdorff type formula}
In this appendix we collect together some results associated with the Baker-Campbell-Hausdorff series used in Section~2.  The basic result which is discussed in many quantum mechanics texts is that
\begin{eqnarray}
  \nonumber\e^A\e^B\e^{-A}&=& B+[A,B]+\frac1{2!}[A,[A,B]]+\frac1{3!}[A,[A,[A,B]]]+\cdots \\
  &=& \e^{\ad{A}}B\,,
\end{eqnarray}
where we define $\ad{A}^nB$ by $\ad{A}^0B=B$, $\ad{A}B=[A,B]$ and $\ad{A}^nB=\ad{A}(\ad{A}^{n-1}B)$. In addition we have the result that if $\delta$ is a derivation then
\begin{equation}\label{bhc2}
\e^{A}\delta\e^{-A}=\frac{1-\e^{\ad{A}}} {\scriptstyle{\ad{A}}}\delta A\,.
\end{equation}
It is helpful to recall how this result is derived as we will need to extend it to the case where we have an additional derivation acting on the whole of this equation.

The proof of~(\ref{bhc2}) rests on the simple identity that
\begin{equation}\label{bhc3}
    \frac{d\ }{ds}\e^{sA}\delta\e^{-sA}=-\e^{sA}(\delta A)\e^{-sA}=-\e^{s\ad{A}}\delta A\,.
\end{equation}
Integrating this we see that
\begin{equation}\label{bhc4}
   \e^{A}\delta\e^{-A} =-\int_0^1 \ud s\, \e^{sA}(\delta A)\e^{-sA}=-\int_0^1 \ud s\,\e^{s\ad{A}}\delta A \,,
\end{equation}
from which the result follows.

We now extend this result and introduce another derivation $\partial$ and calculate $\partial(\e^{A}\delta\e^{-A})$.
Using the Leibnitz property of the derivation, we get three terms, two of which involve the derivation acting on $\e^A$. Using \rf{bhc2} and \rf{bhc4} we find that
\begin{equation}\label{bhc6}
    \partial(\e^{A}\delta\e^{-A})=\frac{1-\e^{\ad{A}}}{\scriptstyle{\ad{A}}}\partial\delta A
    +\left[\frac{1-\e^{\ad{A}}}{\scriptstyle{\ad{A}}},\ad{\frac1{\scriptstyle{\ad{A}}}\partial A}\right]\delta A\,.
\end{equation}
Expanding in powers of $A$ is straightforward and to quadratic accuracy we find that
\begin{equation}\label{bhc7}
    \partial(\e^{A}\delta\e^{-A})=-\partial\delta A-\tfrac12[A,\partial\delta A]-\tfrac12[\partial A,\delta A]\,,
\end{equation}
which is used in the main text.

\section{Open example}
In this appendix we construct an `open' type gauge transformation which obeys $U(0)=\mathbbm{1}$ and $U(x)\rightarrow -\mathbbm{1}$ at spatial infinity. To simplify the construction we will choose $g=1$ and look for an $\alpha_o$ which is monotonically increasing between $0$ and $\pi$ and therefore takes the problematic value $\pi/2$ once and only once.  We demand that $\alpha_o(r)$ has continuous second derivatives in order for $f(r)$ to be continuous, which implies we require three degrees of freedom at each patch. We will construct such an $\alpha_o$ by patching continuous functions together, our ansatz for which will be:
\begin{equation}\label{patch-def}
  \alpha_o(r)=\left\{\begin{array}{cc}
                    Ar e^{-br} & 0\leq r\leq \dfrac{1}{2}, \\
                    \\
                    \dfrac{\pi}{2}+\lambda\big(r-\dfrac{1}{r^2}) & \dfrac{1}{2}\leq r \leq 4, \\
                    \\
                    \pi + \dfrac{d_2}{r^2} + \dfrac{d_3}{r^3}+ \dfrac{d_4}{r^4} & 4\leq r < \infty.
                  \end{array}
            \right.
\end{equation}
We have chosen to patch the functions together at $r=1/2$ and $r=4$, choosing $\alpha_o(1)=\pi/2$ -- note that although the functions chosen are not monotonically increasing on their own they may be made so on their patches for suitable coefficients. Note the choice of the second function, determined by (\ref{behave}).

At $r=1/2$ the equations imposing continuity of $\alpha_o(r)$, $\alpha_o'(r)$, $\alpha_o''(r)$ may be solved to give
\begin{equation}\begin{split}
  A &= \frac{\pi}{96}\big(-121+7\sqrt{865}\big)e^{41/17-\sqrt{865}/17} \simeq 5.492, \\
  b &= \frac{82}{17}-\frac{2}{17}\sqrt{865}\simeq 1.363, \\
  \lambda &= \frac{31\pi}{96}-\frac{\pi}{96}\sqrt{865}\simeq 0.052.
\end{split}\end{equation}
The first patching function has a turning point at $1/b\simeq 0.733$ which is outside its range of definition and is therefore monotonically increasing. The second patching function is always increasing since $\lambda>0$. Continuity at $r=4$ implies
\begin{equation}\begin{split}
  d_2 &= \frac{5067\pi}{32}-\frac{213\pi}{32}\sqrt{865} \simeq -117.566, \\
  d_3 &= -984\pi + 40\pi\sqrt{865}\simeq 604.553, \\
  d_4 &= 1600\pi - 64\pi\sqrt{865}\simeq -886.861.
\end{split}\end{equation}
The derivative of the third patching function has no real roots in its range and is monotonically increasing there. A plot of $\alpha_o(r)$ is given in Figure~\ref{centrea}.

We now turn to the $L^2-$ norm of our configurations. $f(r)$ is a continuous function and is therefore integrable. For large $r$ (i.e. using the third patching function) $f(r)$ may be expanded
\begin{equation*}
  f(r) = 1+ 2\frac{d_3}{d_2}r^{-1} + \mathcal{O}\big(r^{-2}\big).
\end{equation*}
\FIGURE{\centering
\includegraphics[width=0.5\textwidth]{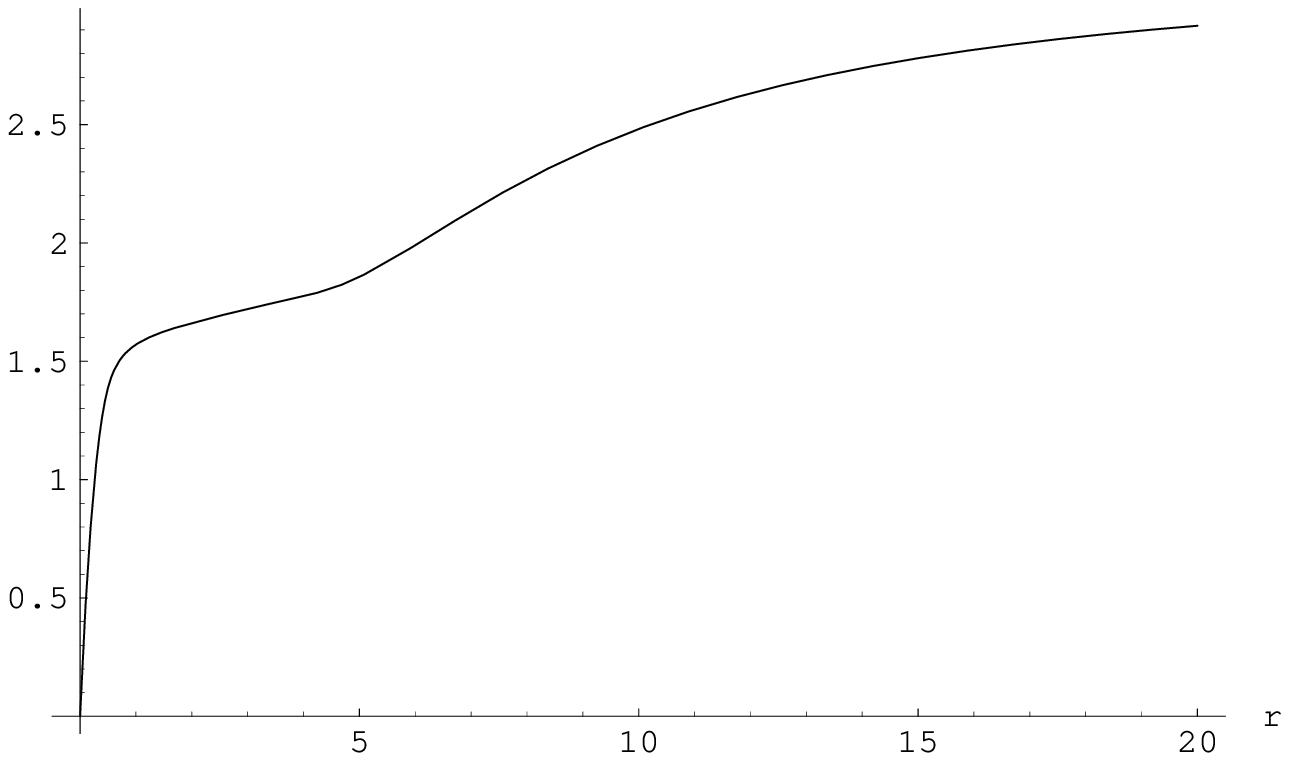}
\label{centrea}\caption{$\alpha_o(r)$ which corresponds to the gauge transformations approaching $-\mathbbm{1}$ at spatial infinity.}}
The integrand of $\|A\|^2$ is therefore bound by $r^{-2}$ at large $r$ and $A$ has finite norm. The copy has norm $\| A^U \|$ given by (\ref{copynorm}). The behaviour of the copy is not severely modified at infinity; $\cos(2\alpha_o)\rightarrow 1$ with $r^{-4}$ corrections which can only improve convergence of the first term. The second and third term go to zero like $r^{-4}$ and the integral will converge. The copy therefore also has finite norm.

The energy is also finite as we now show. The derivative $f'(r)$ is continuous except at the patching points where it has only finite jumps (in particular it is smooth at $r=1$, $\alpha_o=\pi/2$) and is therefore integrable. With reference to (\ref{enorm}) $f(r)$ behaves as follows at the boundaries,
\begin{equation*}\begin{split}
  (f^2(r)-1)^2 &= 16b^2r^2 + \mathcal{O}(r^3)\text{ for $r$ small,} \\
  (f^2(r)-1)^2 &= 16\frac{d_3^2}{d_2^2}r^{-2} + \mathcal{O}(r^{-3})\text{ for $r$ large,} \\
  f'(r) &= -2\frac{d_3}{d_2}r^{-2} +\mathcal{O}(r^{-3}) \text{ for $r$ large,}
\end{split}\end{equation*}
and therefore $\|F\|^2$ is also finite. The copy has the same energy.

Although we have not given a closed form expression for $\alpha_o(r)$ we may make this function and therefore $f(r)$ arbitrarily smooth at the patching points by requiring continuity of the higher derivatives of $\alpha_o$. This is achieved through the inclusion of corrections to the patching functions. Requiring $\alpha_o$ to be monotonically increasing is more restrictive, but this may be relaxed anywhere except very near $\alpha_o=\pi/2$ (or multiples thereof) without further complication.
\section{Axially symmetric copies}
We review Henyey's derivation~\cite{Henyey:1978qd} of the equations for the existence of copies of radially symmetric gauge fields, here including the coupling. The gauge field and gauge transformation are
\begin{equation}
  \mathbf{A} = i a(r,\theta)\hat{\phi}\ \sigma^3=-2a(r,\theta)\hat{\phi}\ T^3,\qquad U=\cos(g\alpha(r))\mathbbm{1}+i\sin(g\alpha(r))\hat{k}\cdot\sigma,
\end{equation}
with $\hat{k}=(\cos(\phi),\sin(\phi),0)$ and $\alpha(r,\theta)=rb(r)\sin(\theta)$. The equation to be solved, c.f. \rf{copies}, is
\begin{equation}
  a(r,\theta) = \frac{1}{2gr\sin(\theta)}-\frac{b(r)+r^2\sin^2(\theta)\bigg(b''(r)+\frac{4}{r}b'(r)\bigg)}{\beta^{-1}\sin\big(2g r\beta b(r)\sin(\theta)\big)}.
\end{equation}
Here $\beta$ is the scaling parameter, $\alpha\to\beta\alpha$ which causes copies to coalesce. The boundary conditions are that $b(r)=b_0+b_2 r^2+\ldots$ with $b_0\not=0$ near the origin and $b(r)$ behaves as $1/r^3$ at infinity. We take Henyey's example,
\begin{equation}
  b(r)=\frac{K}{(r^2+1)^\frac{3}{2}}.
\end{equation}
In the limit $\beta\to0$ the two copies coalesce to the configuration
\begin{equation}
  a(r,\theta)\to \frac{15}{2g}\frac{r\sin(\theta)}{(r^2+1)^2}.
\end{equation}
The non-perturbative nature of this infinitesimal copy is manifest. As proven in \cite{vanBaal:1991zw} this configuration lies on the Gribov horizon.


\end{document}